\documentclass[preprint,proceedings]{rmaa}

\newcommand{\D}{\discretionary{}{}{}}

\SetYear{2013}
\SetConfTitle{XIV Latin American Regional IAU Meeting}

\title{MLS110213:022733+130617: A new eclipsing polar above the period gap}

\author{
  K.M.G. Silva\altaffilmark{1}, C.V. Rodrigues\altaffilmark{2}, A.S. Oliveira\altaffilmark{1}, L.A. Almeida\altaffilmark{3}, D. Cieslinski\altaffilmark{2}, J.E.R. Costa\altaffilmark{2} and F.~J.~Jablonski\altaffilmark{2}}

\altaffiltext{1}{IP\&D,  Universidade do Vale do Para\'{i}ba, Brazil
  (karleyne\D{}@gmail.\D{}com.)}
\altaffiltext{2}{DAS, Instituto Nacional de Pesquisas Espacias, Brazil.}
\altaffiltext{3}{IAG, Universidade de S\~{a}o Paulo, Brazil.}

\suppressfulladdresses

\listofauthors{Silva, K.M.G., Rodrigues C.V., Oliveira, A.S.,Almeida, L.A., Cieslinski, D., Costa, J.E.R, Jablonski, F.J.}
\indexauthor{Silva, K.M.G., Rodrigues C.V., Oliveira, A.S.,Almeida, L.A., Cieslinski, D., Costa, J.E.R, Jablonski, F.J.}

\addkeyword{Stars:Cataclysmic variables}

\begin{document}
\maketitle 
\
Polars are magnetic cataclysmic variables: short-period binaries in which a low-mass main-sequence star transfers matter to a magnetic white dwarf via an accretion column \citep{cropper1990}. A stand-off shock is formed near the white-dwarf surface, followed by a hot post-shock region that cools down by cyclotron radiation mainly in optical and infrared wavelenghs and free-free emission in X-rays. Cyclotron process produces high polarizarion degrees. Over 135 polars and polar candidates were identified so far: 26 percent of them have periods over the period gap ( P \textgreater \ 3h) and only 6 of these systems are eclipsing \citep{ritter2011}. 

We have obtained photometry and polarimetry at Observatorio do Pico dos Dias (OPD) and spectroscopy at SOAR telescope of MLS110213:022733+130617: a polar candidate identified and monitored by the Catalina Sky Survey (CSS, Drake et al. 2009).

\begin{figure}[!t]
  \includegraphics[trim= 3.0cm 1cm 3.0cm 1cm, clip,width=0.84\columnwidth]{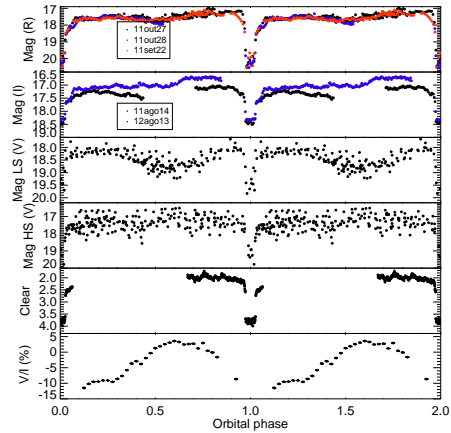}
   \caption{ From top to bottom: R, I, V (low state) and V (high state) bands light curves and R-band circular polarimetric curve of MLS110213.}
  \label{fig:simple}
\end{figure}

Phase-folded photometry and circular polarization are presented in Figure \ref{fig:simple}. From the time of the mid-eclipse observed in the light curves of the OPD photometric data we estimated an orbital period of 3.787 hours. This period is above the period gap and only three other eclipsing systems show longer periods.  CSS long term observations show the system in two distinct brightness states, 1.5 mag apart. This kind of variation in polars is explained as variations in the mass rate accretion. 

Our polarimetric data were obtained in the high state of brightness (HS). The circular polarization varied in the range 3 to -15\% in the R-band, which confirms that this component is mainly due to cyclotron emission from the accretion column. The polarization does not have a standstill at zero, and there are phases when the signal changes, indicating a region always visible, but observed from the backside in some phases, or two accretion regions.

A spectrum in the energy range 4500-7000 \AA \ was obtained convering the phase interval 0.7-0.88. The spectrum shows a flat continuum and no contribution from the secondary star. The spectrum shows the typical characteristic of high ionized optically thick regions: inverse Balmer series decrements (H$_\alpha$~\textless~H$_\beta$~\textless~H$_\gamma$) and intense HeII~$\lambda$4686 are present.

We performed a preliminary modelling of the R-band polarization curves using CYCLOPS code \citep{costa2009,silva2013}. The modeling used one region extended in longitude and located at high latitude. The system orbital inclination provided by the model is 80 degrees, consistent with the observed eclipse. The magnetic field range is 16-33 MG and the mean temperature is 10 keV, both estimated at the emitting region. No X-ray counterpart was identified by ROSAT, RXTE nor Integral X-ray missions in the field observed of MLS110213, indicating a flux lower than the sensibility of those missions.

In summary, our data confirm the system as a new eclipsing polar. Some system properties were estimated.

\end{document}